\documentclass[review]{elsarticle}

\usepackage{hyperref}

\journal{Journal of \LaTeX\ Templates}

\begin{document}

\begin{frontmatter}

\title{A systematic study of the ground state properties of W, Os and Pt isotopes using HFB theory}

\author{NITHU ASHOK\corref{nit}}
\cortext[nit]{Corresponding author}
\ead[e-mail]{nithu.ashok@gmail.com}
\author{ANTONY JOSEPH}
\address{Department of Physics, University of Calicut,\\
 Kerala, India}




\begin{abstract}
A systematic study of the ground state properties of transitional nuclei W, Os and Pt is conducted with the help of Skyrme Hartree-Fock-Bogoliubov theory.  Different Skyrme interactions are employed in the study.  Two different bases, harmonic oscillator and transformed harmonic oscillator, are used in our investigation.  2n- separation energy, charge radii, neutron and proton rms radii, neutron skin thickness and deformation parameter have been estimated.  The results obtained are in good agreement with the available experimental values.

\end{abstract}

\begin{keyword}
\texttt{Hartree-Fock-Bogoliubov, binding energy, neutron separation energy, rms radii.}
\end{keyword}

\end{frontmatter}


\section{Introduction}

The developments of experimental facilities like Radioactive Ion Beams (RIB) help us to study the
structural properties of a wide range of nuclei in the nuclear chart\cite{rib,rib1}. 
Experimental facilities which exist today are incapable of studying nuclei very far from the stability
line. To explore their properties, we have to rely mainly on theoretical methods.  Many theoretical studies have been 
carried out in recent years by various groups to study their structural properties.  Investigation of nuclear structure towards drip-line has become a hot research topic nowadays. 
Nuclei near the drip-line, owing to the low binding, exhibit several interesting phenomena like neutron skin, halo etc.  
Experimental evidence for neutron skin in $^{208}\textrm{Pb}$ have been observed by various methods \cite{pb1,pb2,pb3,pb4,pb5} in recent years.  Theoretically, with the help of
 various phenomenological as well as microscopic models, neutron skin thickness have been predicted for a wide range of nuclei\cite{th1,th2,th3,th4}.
Moreover, neutron rich nuclei near to drip-line have great relevance in stellar nucleosynthesis \cite{kur,kaj}.

Transitional nuclei serves an interesting region for the study of the nuclear structural properties because of the presence of the oblate shaped isotopes.   
Several authors have analysed the structural properties and shape evolution of transitional nuclei, Yb, Hf, W, Os, Pt etc using various experimental \cite{alkh,whel,john,pt1,pt2} as well as theoretical \cite{ans,znaik,stev,sar2008,rob2009,rod2010,rmf}models in recent decades.
Nuclear size is one of the important characteristic properties of a nucleus.  Nuclear mass and radii are the main quantities used to probe the structure of a nucleus.

In the present study, we have tried to study the structural properties of even-even W, Os and Pt isotopes using Skyrme-Hartree Fock-Bogoliubov (HFB)theory.  In our earlier studies, we have successfully applied this theory in predicting the decay properties of these nuclei\cite{n1,n2,n3}.  This theory serves well in the region away from drip line due to the inclusion of pairing correlation and thereby removing the continuum problems\cite{cont,cont1}.  Moreover, we have used two different basis to solve HFB equation, Harmonic oscillator (HO) and Transformed harmonic oscillator (THO).  HO basis explains the nuclear properties in the nuclear interior in the case of nuclei near and far from beta stability line, very well.  But in the case of the exterior part of the drip line or weakly bound nuclei, HO basis expansion converges slowly and results in the reduction of densities and do not reflect the pairing correlations correctly.  However, the THO basis, explains the exterior part as well.

Nuclei in the region $A\sim 190$ are intrinsically deformed in their ground state itself.
These nuclei belong to the transitional region where a shape change between prolate, oblate and spherical configuration is observed. Certain studies showed that nuclei in these regions are superdeformed. These
properties make them interesting candidates for structure studies.  As a representative of this region, we have  selected Tungsten (W), 
Osmium (Os) and Platinum (Pt) isotopes for our investigation.   Moreover, these nuclei are near to the doubly magic $^{208}\textrm{Pb}$ nucleus.

The paper is organised as follows.  In section \ref{sec2}, we have given the theoretical formalism, along with the details of calculations 
employed in the present work.  The results of the calculations and their relevant discussion are given in section \ref{sec3}.  
Finally, the concluding remark is presented in section \ref{sec4}.

\begin{figure*}[h]
\centerline{\includegraphics[width=7in]{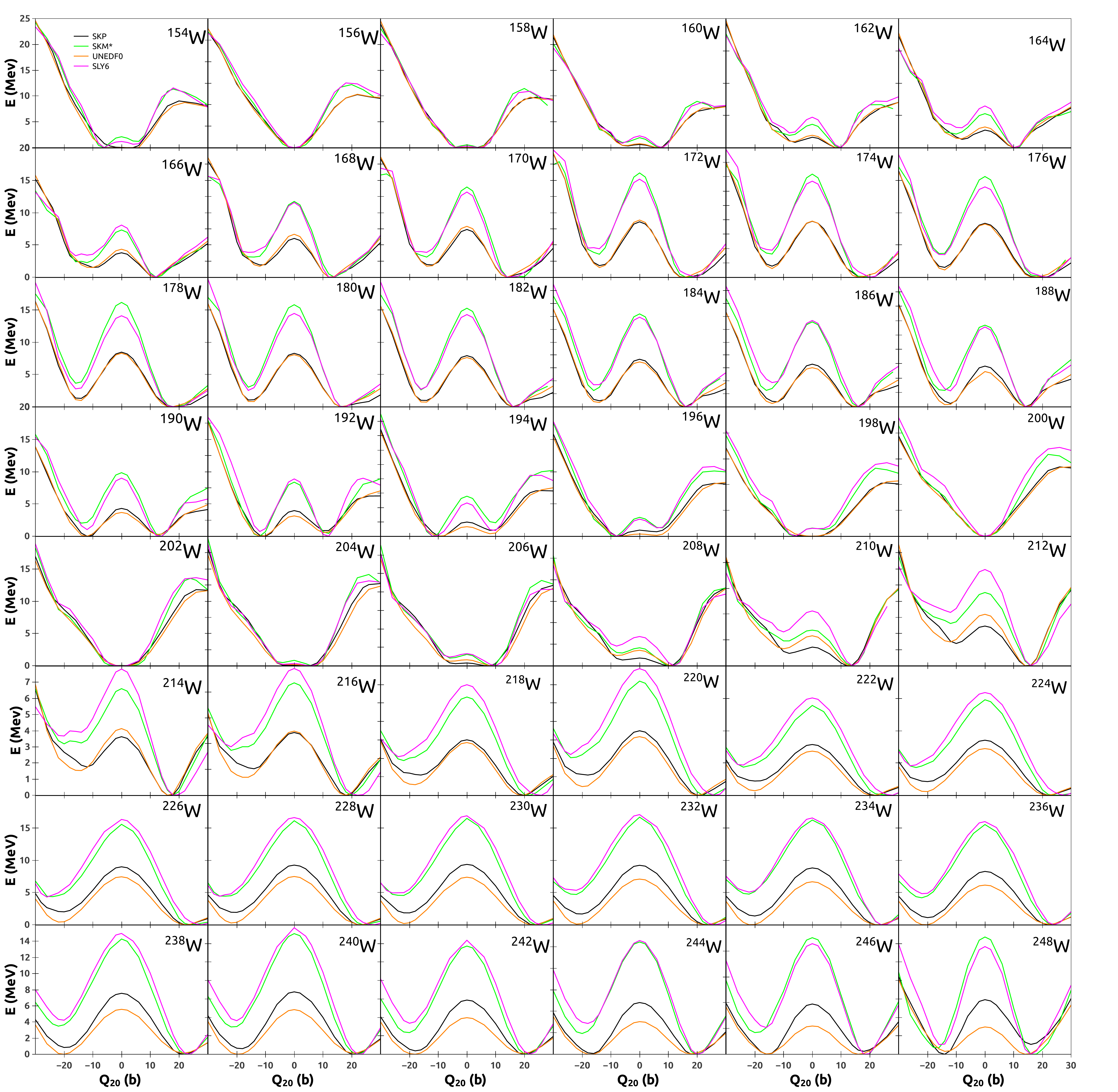}}
\vspace*{8pt}
\caption{Potential energy curves of W isotopes for some selected Skyrme forces.}
\protect\label{pec}
\end{figure*}
  
\section{Theoretical framework}\label{sec2}
  Hartree-Fock-Bogoliubov (HFB) theory, which is the generalised form of HF+BCS theory, consider both the long range mean field part and
the short range pairing part with equal importance\cite{ring}.     
  In the matrix form, the HFB equation is given by\cite{bend},
    \begin{equation}
     \left( {\begin{array}{cc}
h-\lambda   &  \Delta   \\
-\Delta ^{*}      &   -h^{*}+\lambda 
\end{array} }\right)
\left( {\begin{array}{c}
U_{n}   \\
 V_{n}
\end{array} }\right)
=E_{n}\left( {\begin{array}{c}
U_{n}   \\
 V_{n}
\end{array} }\right)
     \end{equation}     
where $h=t+\Gamma $ is the HF potential, $\Delta $ is the pairing potential, $E_{n}$ is the quasiparticle energy, $\lambda $ is the 
chemical potential and $U_{n}$ and $V_{n}$ are the upper and lower components of the quasiparticle wavefunction.

Skyrme HFB equations are solved using axially deformed cylindrical harmonic oscillator (HO) and transformed harmonic oscillator (THO)
 basis\cite{hfbtho}.  Numerical calculations have been done with the help of 20 oscillator shells with a cut-off energy of 60 MeV.  
In the mean field part, we used the zero range Skyrme effective interaction.  There exists a variety of Skyrme parametrizations.  
In the present work, we made use of some widely used Skyrme interactions like SIII\cite{siii}, SKP\cite{skp}, SLY6\cite{sly6}, SKM*\cite{skm}, 
UNEDF0\cite{une0} and UNEDF1\cite{une1}. These Skyrme forces are found to be very efficient in reproducing nuclear ground state properties.  
In the pairing part, density dependent delta interaction (DDDI)\cite{dddi,dddi1} in the mixed form is used.  It is expressed as \cite{mix},
\begin{equation}
  V^{n/p}_{\delta }(\vec{r_{1}},\vec{r_{2}})=V_{0}^{n/p}[1-\frac{1}{2}(\frac{\rho(\vec{r_{1}} +\vec{r_{2}})}{\rho _{0}} )^{\alpha }]\delta (\vec{r_{1}}-\vec{r_{2}})
\end{equation}
where the saturation density\cite{sat} $\rho _{0}$=0.16 fm$^{-3}$ and $\alpha $=1.
\begin{figure}
\centerline{\includegraphics[width=4in]{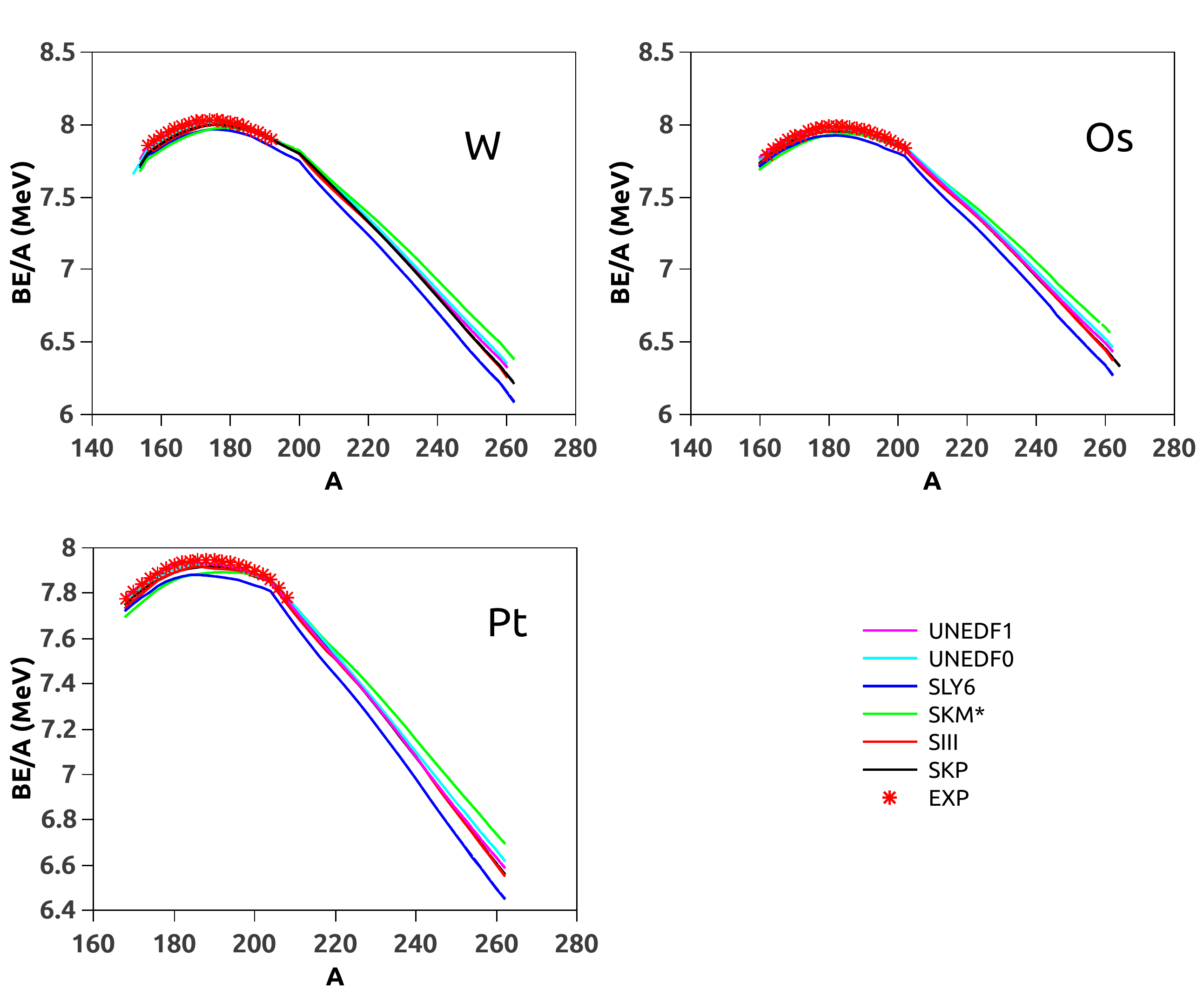}}
\vspace*{8pt}
\caption{Be/A for W, Os and Pt isotopes calculated using HO (solid) and THO (dashed) basis.}
\protect\label{bea}
\end{figure}
The study had begun with the search for drip-line nuclei where the nuclear binding ends. Separation energy is the energy required to remove 
the last nucleon from the nucleus.  At the drip line, the separation energy should approach zero.  
Here, we are interested in the 2n-separation energy.  It is estimated from the binding energy using the relation,
\begin{equation}\label{eq:s2n}
S_{2n}(N,Z)=BE(N,Z)-BE(N-2,Z)
\end{equation}

The quantity which specifies the shape of a nucleus is the quadrupole moment.  From the quadrupole moment, we have estimated the deformation parameter $ \beta _{2} $ using the expression,
\begin{equation}
  \beta _{2}=\frac{4\pi }{3R^{2}A}\sqrt{\frac{5}{16\pi }}Q
  \end{equation}  
where Q is the quadrupole moment, A is the mass number and $ R=R_{0} A^{1/3} $, with $ R_{0} =1.2$ fm.

We are also interested in the investigation of neutron and proton distribution in W, Os and Pt isotopes.  The quantity which characterizes these are the neutron and proton radii.
The mean square radius of neutron and proton can be obtained from nucleonic density ($ \rho _{p,n} $)\cite{rad} and are given by
\begin{equation}
 < r^{2}_{p,n}>= \frac{\int R^{2}\rho _{p,n}(R)d^{3}R}{\int \rho_{p,n}(R)d^{3}R }
\end{equation}
and finally rms radii is given by 
\begin{equation}
  r^{p,n}_{rms}=\sqrt{< r^{2}_{p,n}>}
  \end{equation}
  Nuclear charge radii are obtained by folding the proton distribution with finite size of neutron and proton.
  Nuclear rms charge radius is evaluated using the simplified form of the expression \cite{bend}
  \begin{equation}
  r_{c}=\sqrt{r_{p}^{2}+0.64}
  \end{equation} 
 \begin{figure}
\centerline{\includegraphics[width=4in]{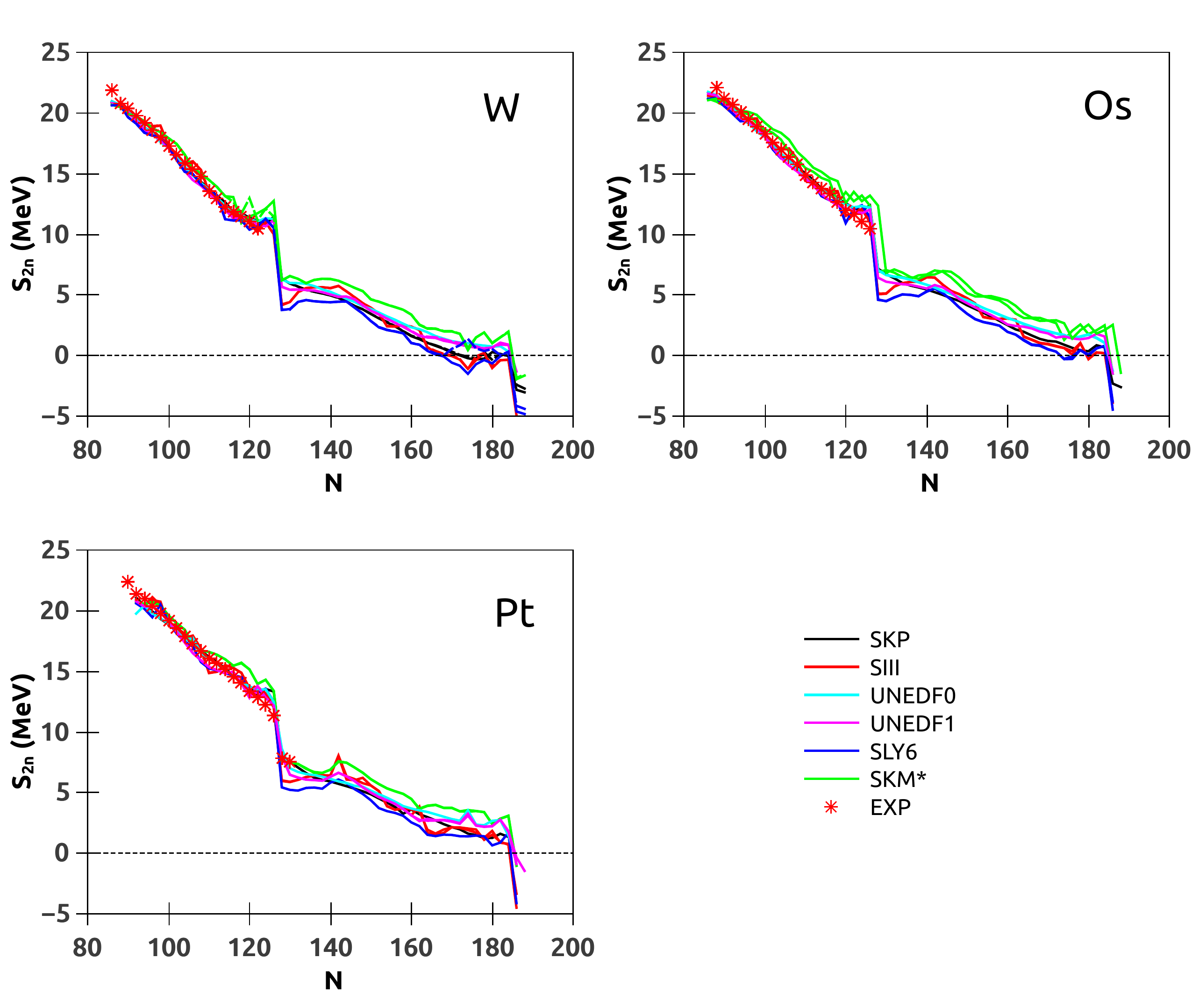}}%
\vspace*{8pt}
\caption{2n-separation energy calculated using HO (solid) and THO (dashed) basis.}
\protect\label{s2n}
\end{figure} 
  For the nuclei far from the stability line, the number of neutrons is much greater than that of the protons.  This leads to the formation of a thin layer of neutrons around the bulk matter of the nucleus.  This layer which is termed as the 
neutron skin is measured as the difference between the neutron and proton rms radii \cite{miz}.
  \begin{equation}
skin\  thickness\equiv <r^{2}_{n}>^{1/2}-<r^{2}_{p}>^{1/2}
\end{equation}
\begin{figure}
\centerline{\includegraphics[width=4in]{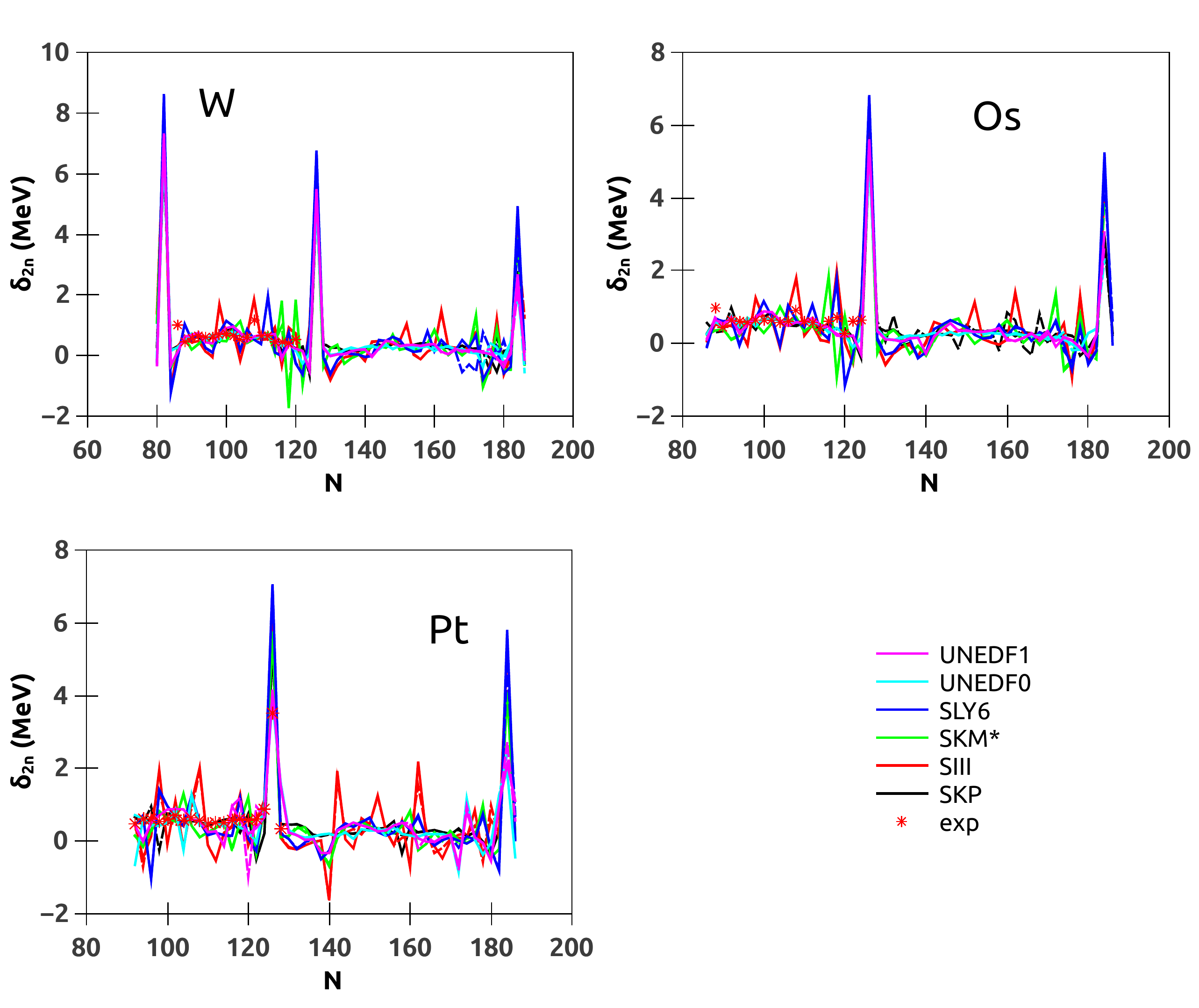}}
\caption{2n-shell gap calculated using HO (solid) and THO (dashed) basis.}
\label{shlg}
\end{figure} 
\begin{figure}
\centerline{\includegraphics[width=4in]{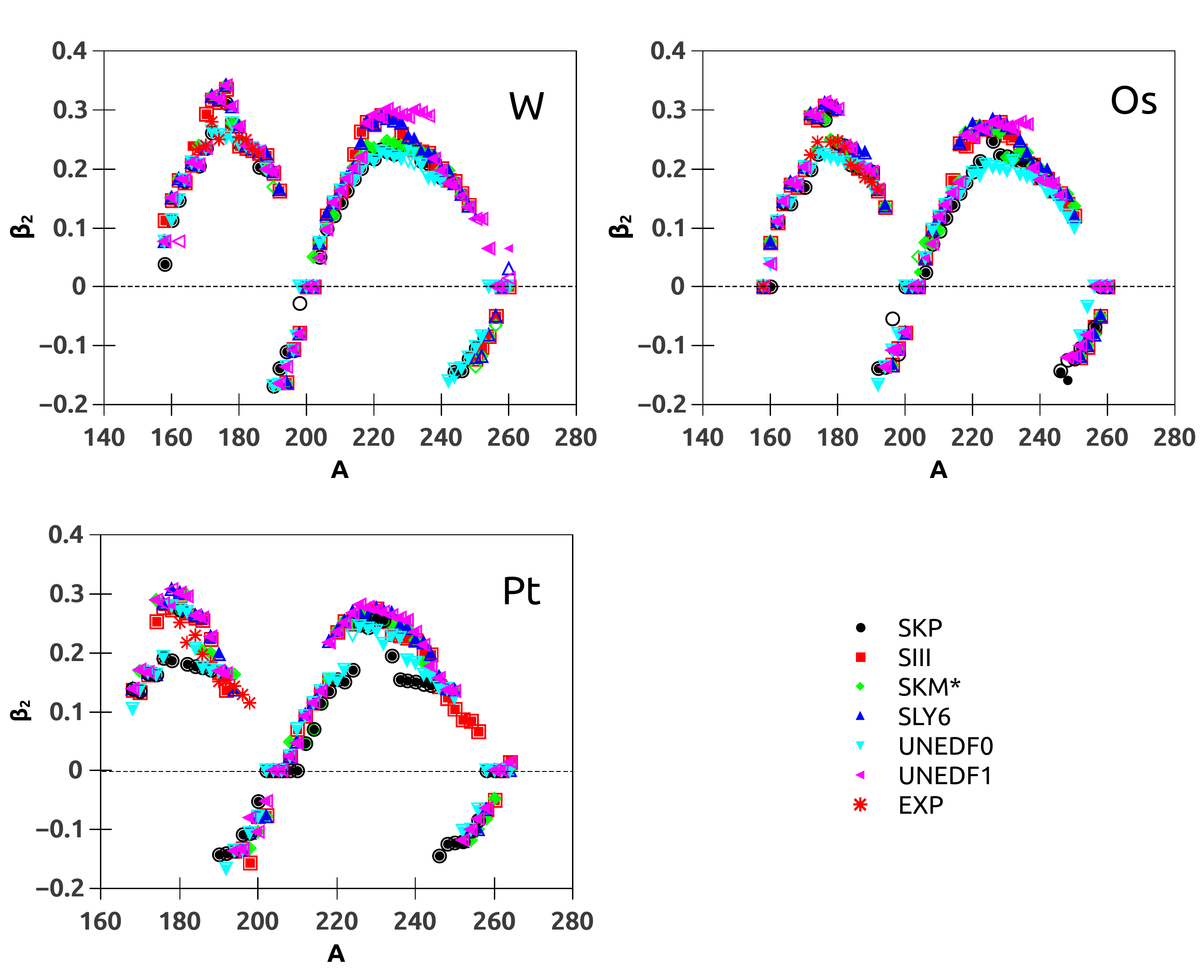}}
\caption{Deformation parameter calculated using HO (solid) and THO (dashed) basis.}
\label{def-full}
\end{figure} 
\section{Results and discussion}\label{sec3}
In the present work, we have made an attempt to study systematically the ground state properties of even-even W, Os and Pt isotopes.  The selected isotopes (W, Os and Pt) ranges from 2p drip-line to 2n drip-line.  The subsections deals with the study of the bulk properties of these isotopes.
\subsection{Potential Energy Curves}
We have started by developing the  Potential Energy Curves (PEC) by using the linear constrained method\cite{constr}.  In this method, we have to minimize $ E^{'}=E-\lambda (\hat{<Q_{20}}>-Q_{20}) $, where $\lambda $ is the Lagrange multiplier and $Q_{20} $ is the quadrupole moment.  Thus, binding energy corresponding to a specific $Q_{20} $ is obtained.  Total energy which corresponds to the minimum value is the ground state and all other local minima are the excited intrinsic states.  We are interested only in the ground state properties.  Fig. \ref{pec} shows the PEC of W isotopes corresponding to HO basis for some selected Skyrme forces.  Similar plots are also there for Os and Pt.  Since the PECs follow a similar trend in the case of Os and Pt isotopes, we have not shown them here.  From the PECs we can see that the shape of the nuclei systematically changes betweeen prolate, oblate and spherical configurations. 
\subsection{Binding energy}
Binding energy is one of the important quantities which help to analyse the validity of a theoretical model.  The ground state binding energies per nucleon are evaluated and plotted in Fig. \ref{bea}.  The results are compared with the available experimental values \cite{ame2016}.  The calculated binding energies per nucleon shows very small deviation from the experimental values.  Compared to other Skyrme parameters, UNEDF values are in better agrement with experimental values.  The parabolic shape of the graph is also reproduced in the case of all Skyrme forces.

\subsection{Separation energies and deformation parameter}
\begin{figure}
\centerline{
\includegraphics[width=4in]{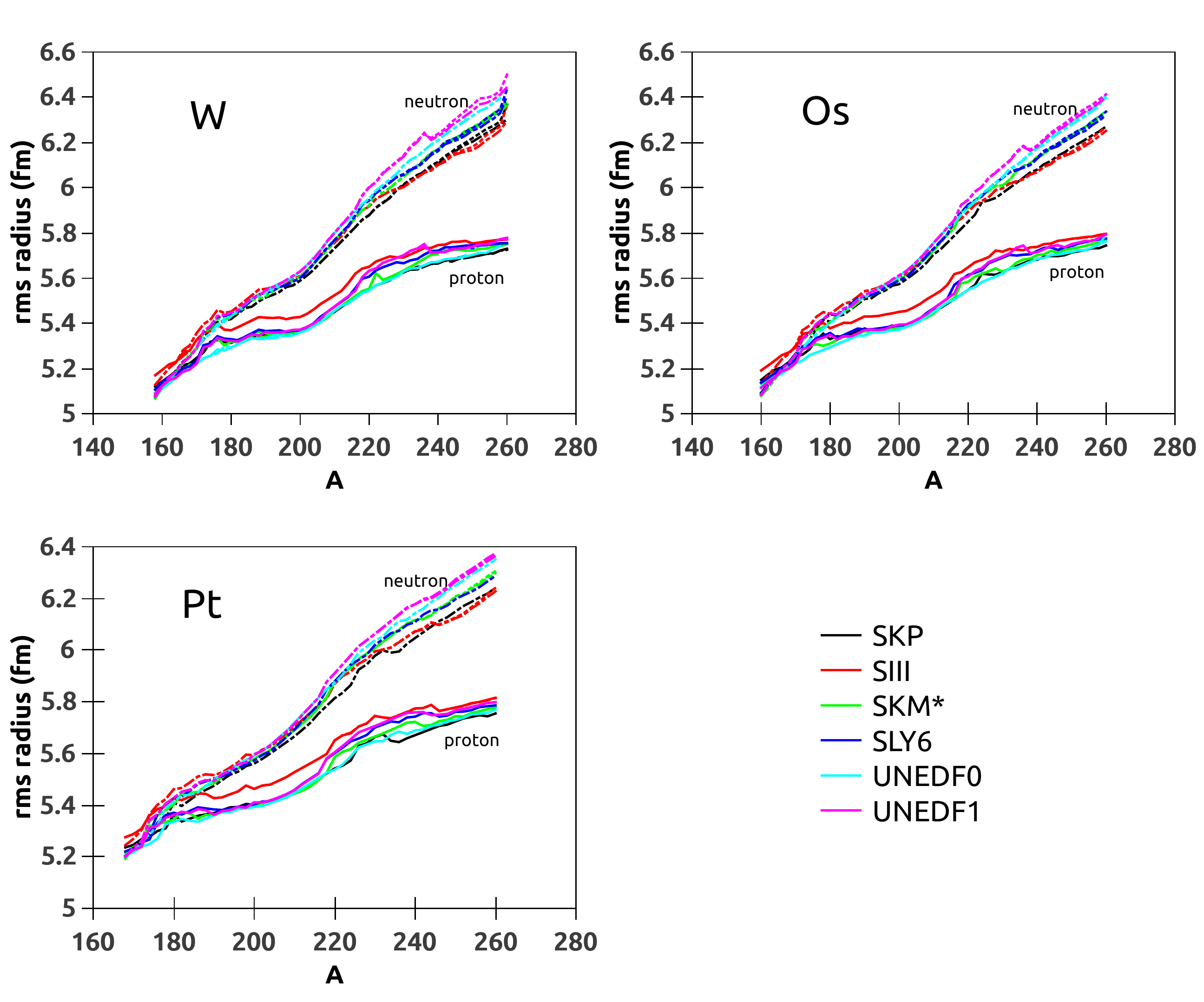}}%
\caption{Neutron and proton rms radii calculated using HO and THO basis }
\label{rms}
\end{figure}
In fig \ref{s2n}. we have shown the 2n-separation energies which are evaluated from the binding energies using equation (\ref{eq:s2n}).

  The calculated results have been compared with the available experimental values \cite{ame2016}.
From the figure, we can see that, as the neutron number increases, ie, on aproaching towards the drip line, the separation energy decreases.  This is 
evident from the fact that as the number of neutrons increases, the nucleons becomes less bound with each other and a small amount of energy is required 
to remove them from the nucleus.  From the figure, a sudden fall in the separation energy is observed at the neutron shell closure (N=126).  
For all the three nuclei, again a fall in the separation energy is observed at N=184.  We can assume that at this neutron number the next shell 
closure exits and can be the next magic number after 126.  We can see similar observations reported ealier using RMF calculations in the case of superheavy 
nuclei \cite{magic}.

In order to corfirm the shell closure, we have evaluated 2n-shell gap, which is the differential variation of 2n-separation energy.  It is expressed as,
\begin{equation}
  \delta _{2n}=S_{2n}(N,Z)-S_{2n}(N+2,Z)
\end{equation}
The computed $\delta _{2n}$ is plotted in Fig. \ref{shlg}.  From the figure, we can observe a sharp peak at N = 82, 126 and 184 in the case of W isotopes.  For Os and Pt isotopes, sharp peak is observed for N = 126 and 184.  At N = 82 also we will observe the peak, which is not shown here, as it is outside the selected range of isotopes.  $\delta _{2n}$ values also confirms the magic character of N=184.   

We have also estimated the deformation parameters of these nuclei.  The computed values are depicted in fig. \ref{def-full}.  It shows that at and near to the magic numbers (N=126 and 184), the deformation parameter is zero, showing spherical configuration.  We can observe the change in the deformation parameter from prolate to oblate and then to spherical configuration.
\subsection{Nuclear radii}
\begin{figure}
\centerline{\includegraphics[width=4in]{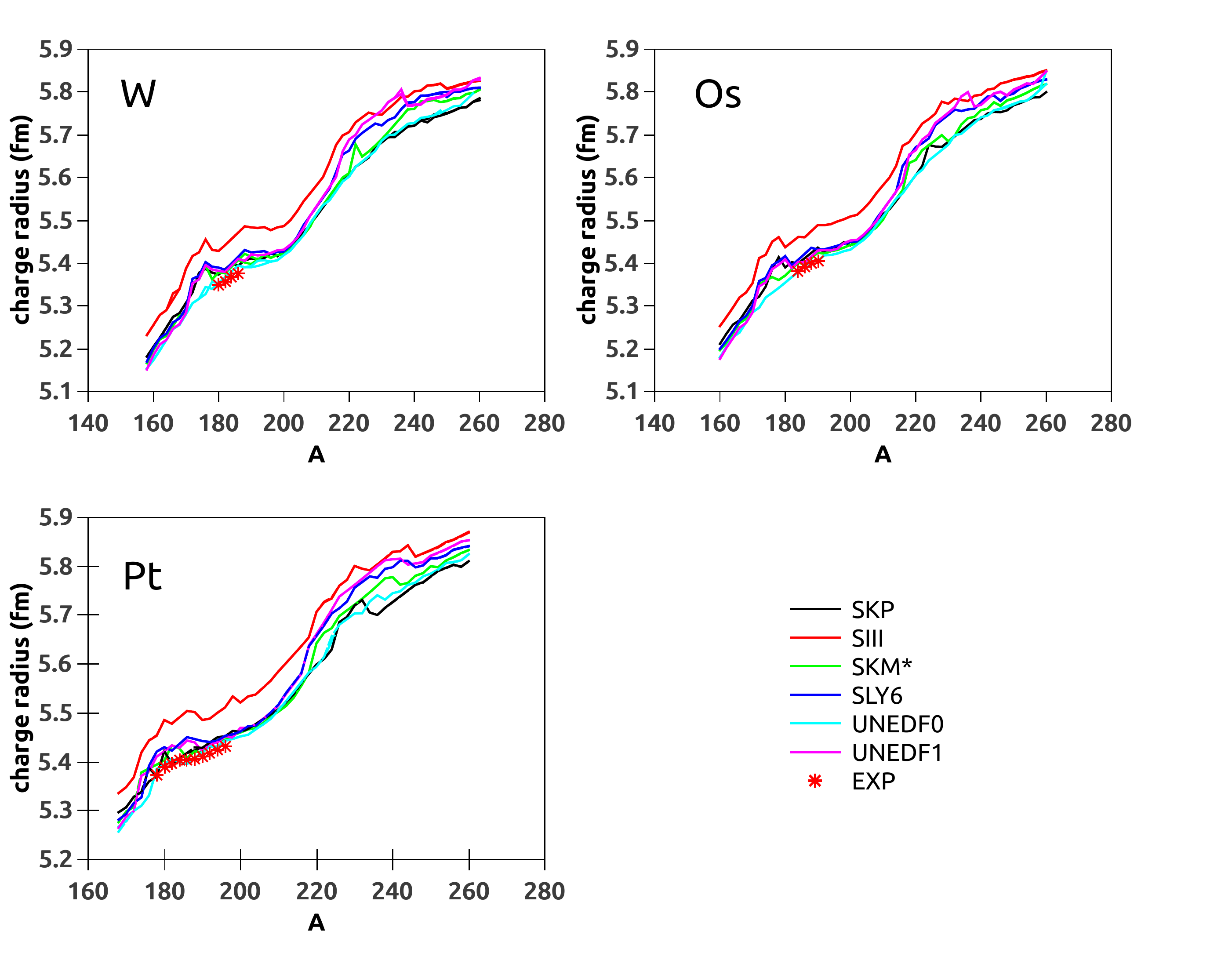}}%
\caption{Nuclear charge radii calculated using HO (solid) and THO (dashed) basis}
\label{charge}
\end{figure} 
Nuclear radius is one of the important quantities which helps in studying the structural properties of nuclei.  Fig. \ref{rms} 
shows the rms radii of proton and neutron distributions computed using different Skyrme forces.  As we move away from 
the beta stability line, it is observed that the 2n-separation energy decreases and finally goes to zero at the drip line. 
 This reduction in the 2n-separation energy with neutron number results in the spatial extension of neutron distribution. 
 This means that neutrons in the nuclei near to drip-lines are weakly bound.  For estimating this spatial extension, we use the 
quantity rms radii.  From the figure, we can visualise that as the mass number increases, neutron rms radii increases.  
Even if the proton number is a constant for an isotope, due to the n-p interaction, the proton radius increases.  
 
Fig. \ref{charge} shows the nuclear charge radii.  We have compared them with the available experimental data.  
All the Skyrme forces shows similar trend in predicting the nuclear charge radii.  SIII values overstimate the predicted charge radii. All 
the other forces predict more or less same charge radii.
\subsection{Neutron skin}
\begin{figure}
\centerline{\includegraphics[width=4in]{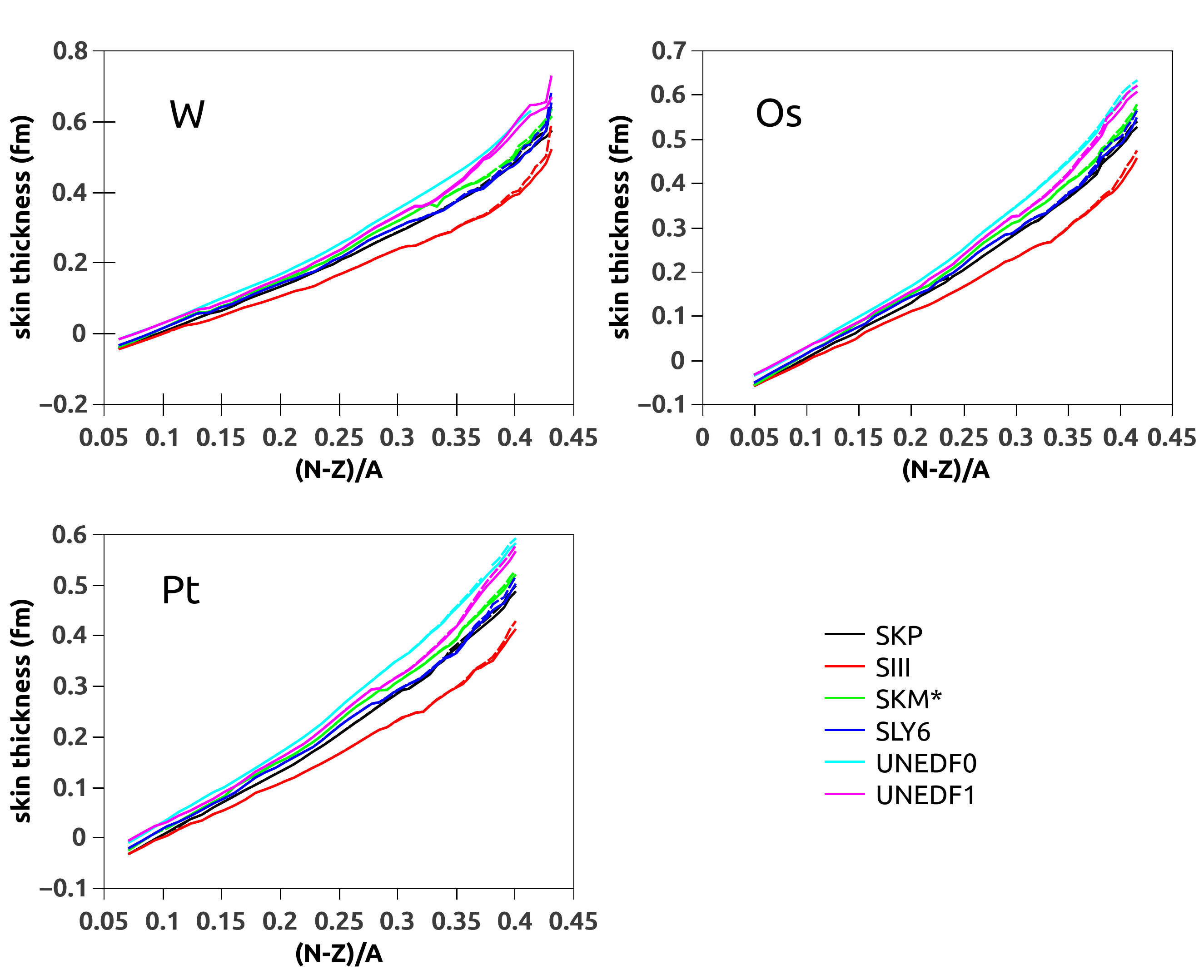}}%
\caption{Neutron skin thickness calculated using HO (solid) and THO (dashed) basis }
\label{skin}
\end{figure} 
Another interesting phenomenon is the formation of neutron skin in the nuclei near drip-line.  Because of the spatial extension of neutrons 
around the nuclear core, a thin layer of neutrons is expected to evolve, which we call the neutron skin.  This layer of neutrons is quantitatively 
expressed as the difference between neutron and proton rms radii.  For normal nuclei, i.e. nuclei near to the beta-stability line this difference are found 
to be 1-2 fm.  But as the neutron number increases towards the drip-line, the radial dimension of neutron distribution is much greater than that of 
protons.  Hence the difference between neutron and proton rms radii also increases.  If it exceeds the above-mentioned value we can expect the formation 
of neutron skin \cite{miz}.     
  In fig. \ref{skin} we have plotted the neutron skin thickness of W, Os and Pt isotopes, as a function of asymmetry parameter, estimated using various Skyrme forces.  From the figure,  
we can see that the thickness of the skin increases with neutron number.  Like our earlier observations, here also, all the Skyrme forces shows 
the same trend in predicting the values of neutron skin thickness.  We obtain a linear relationship between the skin thickness and mass number 
(or neutron number).  It is found that the SIII force underestimates the values of the predicted skin thickness all the other Skyrme forces.
 UNEDF values predict a large skin thickness for all the nuclei under consideration.
 \begin{figure}
\centerline{\includegraphics[width=4in]{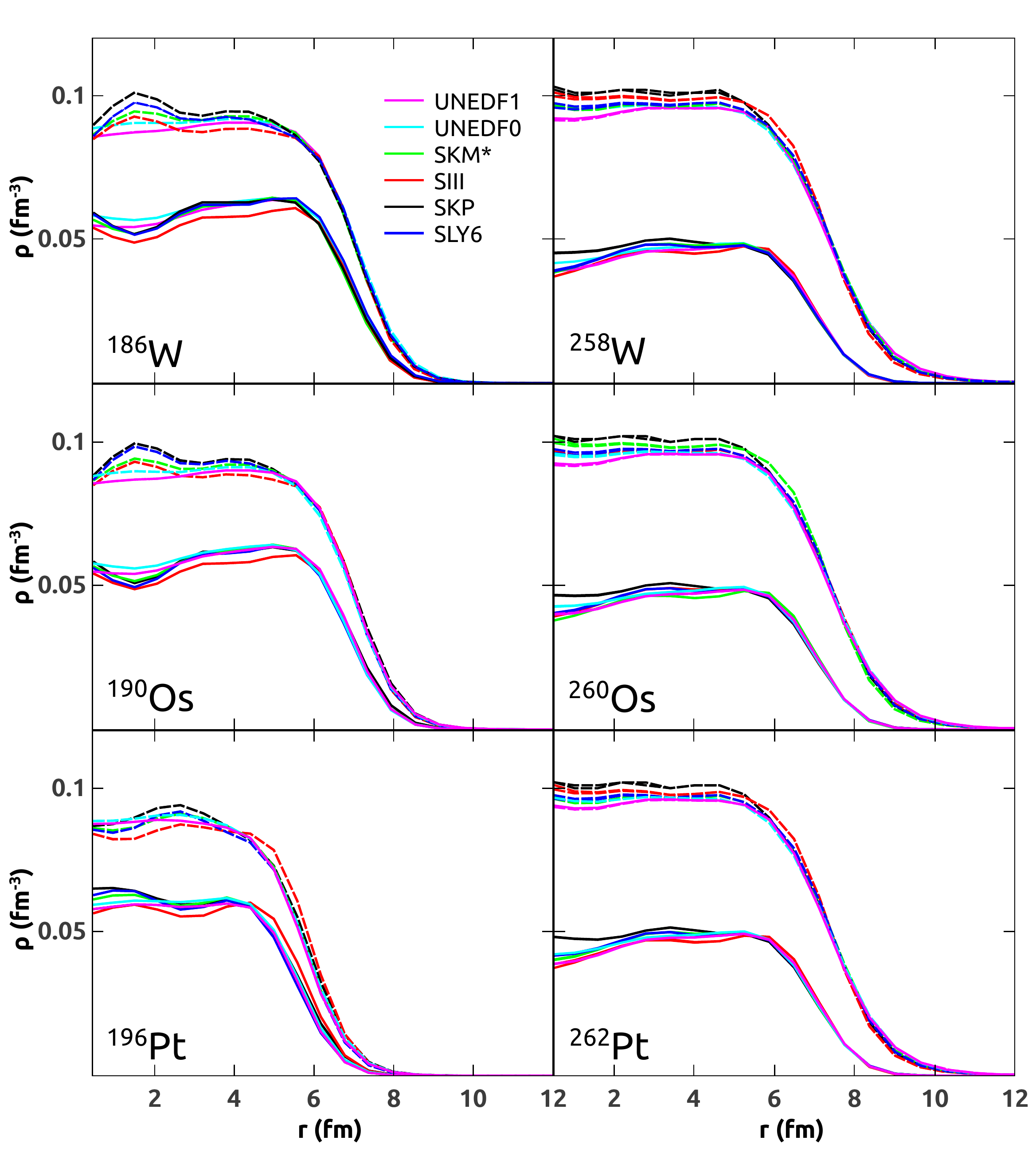}}%
\caption{Neutron and proton density distribution of W (top), Os (middle) and Pt (bottom) using HO (solid) and THO (dashed) basis }
\label{den}
\end{figure}

The evolution of neutron skin can also be visualized with the help of density profile.  We have plotted in fig. \ref{den} the neutron and proton density 
distributions 
of the stable and drip-line isotopes of W, Os and Pt nuclei.  We can predict the formation of neutron skin in these isotopes by analysing 
the distance between the tail of neutron and proton density distributions.  In the case of nuclei near to beta-stability line this distance will be very small.  From the figure, we have observed that as neutron number increases, this distance also increases.  This shows the formation of neutron skin in the isotopes towards the neutron drip-line.

\section{Conclusion}\label{sec4}
In summary, we have made an attempt to investigate some of the structural properties of
transitional nuclei W, Os and Pt.  The study has been carried out with the help of Skyrme HFB theory.  Harmonic oscillator and Transformed harmonic 
oscillator basis have been employed in solving Skyrme HFB equations.    In the present work, we made use of some widely used Skyrme interactions 
like SIII, SKP, SLY6, SKM*, UNEDF0 and UNEDF1.  We have studied the 2n-separation energies, neutron and proton rms radii and neutron and proton density 
distributions in these isotopes.  We have predicted a sudden fall in 2n-separation energy at N=126 and 184.  The calculated deformation parameters 
also show the spherical nature near N=184.  So we have concluded that N=184 may be the next neutron magic number.  We have also observed that as 
the neutron number increases, the neutron and proton rms radii increases.  In W, Os and Pt isotopes, on approaching the neutron drip line, 
the difference between neutron and proton radii increases which results in the formation of neutron skin.  The same has been supported with the 
aid of the density profile.  In short, we have studied, with the aid of different Skyrme forces, the structural properties of nuclei lying in the unexplored regions of the nuclear chart.   

\section*{Acknowledgments}

One of the authors, (NA) gratefully acknowledges UGC, Govt. of India, for providing the grant under UGC-JRF/SRF scheme and also for providing facilities made available under UGC-SAP-DRS II project.


\begin{thebibliography}{0}
\bibitem{rib}J. Dobaczewski and W. Nazarewicz, {\it Philos. Trans. R. Soc. London A}, {\bf 356}, 2007(1998).
\bibitem{rib1}R. F. Casten and B. M. Sherrill, {\it Prog. Part. Nucl. Phys.}, {\bf 45}, S171(2000).
\bibitem{pb1} S. Abrahamyan et al. (PREX Collaboration), {\it Phys. Rev. Lett.}, {\bf 108}, 112502(2012).
\bibitem{pb2} J. Zenihiro et al., {\it Phys. Rev. C}, {\bf 82}, 044611(2010).
\bibitem{pb3} B. Klos et al., {\it Phys. Rev. C}, {\bf 76}, 014311(2007).
\bibitem{pb4} A. Krasznahorkay et al., {\it Nucl. Phys. A}, {\bf 31}, 224(2004).
\bibitem{pb5} C. M. Tarbert et al., {\it Phys. Rev. Lett.}, {\bf 112}, 242502(2014).
\bibitem{th1} W. D. Myers and W. J. Swiatecki, {\it Ann. of Phys.}, {\bf 55}, 395(1969).
\bibitem{th2} M. M. Sharma and P. Ring, {\it Phys. Rev. C}, {\bf 45}, 2514(1992).
\bibitem{th3} N. Fukunishi, T. Otsuka and I. Tanihata, {\it Phys. Rev. C}, {\bf 48}, 1648(1993).
\bibitem{th4} J. Dobaczewski, W. Nazarewicz and T.R. Werner, {\it Z. Phys. A}, {\bf 354}, 27(1996). 
\bibitem{kur} J. Kurcewicz et al., {\it Phys. Lett. B}, {\bf 717}, 371(2012).
\bibitem{kaj} T. Kajino and G. J. Mathews, {\it Rep. Prog. Phys.}, {\bf 80}, 8(2017).
\bibitem{alkh} N. Alkhomashi, et. al, {\it Phys. Rev. C}, {\bf 80}, 064308(2009).
\bibitem{whel} C. Wheldon, et. al, {\it Phys. Rev. C}, {\bf 63}, 011304(R)(2000).
\bibitem{john} P. R. John, {\it Phys. Rev. C}, {\bf 90}, 021301(R)(2014).
\bibitem{pt1} A. Rohilla et. al, {\it Eur. Phys. J. A}, {\bf 53}, 64(2017).
\bibitem{pt2} S. K. Chamoli, {\it Acta Phy. Pol. B}, {\bf 48}, 337(2017).

\bibitem{ans} A. Ansari, {\it Phys. Rev. C}, {\bf 33}, 321(1986).
\bibitem{znaik} Z. Naik, B. K. Sharma, T. K. Jha, P. Arumugam and S. K. Patra, {\it Pramana}, {\bf 62}, 827(2004).
\bibitem{stev} P. D. Stevenson, M. P. Brine, Zs. Podolyak, P. H. Regan, P. M. Walker and J. Rikovska Stone, {\it Phys. Rev. C}, {\bf 72}, 047303(2005).
\bibitem{sar2008} P. Sarriguren, R. Rodriguez-Guzman and L. M. Robledo, {\it Phys. Rev. C}, {\bf 77}, 064322(2008). 
\bibitem{rob2009} L. M. Robledo, R. Rodriguez-Guzman and P. Sarriguren, {\it J. Phys. G: Nucl, Part, Phys}, {\bf 36}, 115104(2009).
\bibitem{rod2010} R. Rodriguez-Guzman, P. Sarriguren, L. M. Robledo and J. E. Garcia-Ramos, {\it Phys. Rev. C}, {\bf 81}, 024310(2010).

\bibitem{rmf} S. Mahapatro, C. Lahiri, B. Kumar, R. N. Mishra and S. K. Patra, {\it Int. J. Mod. Phys. E}, {\bf 25},1650062(2016).
\bibitem{n1} N. Ashok, D. M. Joseph and A. Joseph, {\it Mod. Phys. Lett. A}, {\bf 31}, 1650045(2016).
\bibitem{n2} N. Ashok and A. Joseph, {\it  Nucl. Phys. A}, {\bf 977}, 101(2018).
\bibitem{n3} N. Ashok and A. Joseph, {\it  Int. J. Mod. Phy. E}, {\bf 27}, 1850098(2018).
\bibitem{cont} J. Dobaczewski, W. Nazarewicz, T.R. Werner, J. F. Berger, C. R. Chinn and J. Decharge, {\it Phys. Rev. C}, {\bf 53}, 2809(1996).
\bibitem{cont1} J. Dobaczewski, I. Hamamoto, W. Nazarewicz and J. A. Sheikh, {\it Phys. Rev. Lett.}, {\bf 72}, 981(1994). 
\bibitem{ring} P. Ring and P. Shuck, {\it The Nuclear Many-Body Problem}, (Springer, Berlin, 1980).

\bibitem{bend} M. Bender, P.H. Heenen and P.G. Reinhard, {\it Rev. Mod. Phys}, {\bf 75}, 121(2003). 
\bibitem{hfbtho} M.V. Stoitsov, N. Schunck, M. Kortelainen, N. Michel, H. Nam, E. Olsen, J. Sarich and S. Wild, {\it Comp. Phys. Commun.}, {\bf 184}, 1592(2013).

\bibitem{siii} M. Beiner, H. Flocard, N. Van Giai, and P. Quentin, {\it Nucl. Phys. A.}, {\bf 238}, 29(1975).
\bibitem{skp} J. Dobaczewski, H. Flocard and J. Treiner, {\it Nucl. Phys. A}, {\bf 422}, 103(1984).
\bibitem{sly6} E. Chabanat, P. Bonche, P. Haensel, J. Meyer and R. Schaeffer, {\it Nucl. Phys. A}, {\bf 635}, 231(1998).
\bibitem{skm} J. Bartel, P. Quentin, M. Brack, C. Guet, and H. B. Hakansson, {\it Nucl. Phys. A}, {\bf 386}, 79 (1982).
\bibitem{une0} M. Kortelainen, T. Lesinski, J. More, W. Nazarewicz, J. Sarich, N. Schunck, M.V. Stoitsov and S. Wild, {\it Phys. Rev. C}, {\bf 82}, 024313(2010).
\bibitem{une1} M. Kortelainen, J. McDonnell, W. Nazarewicz, P.G. Reinhard, J. Sarich, N. Schunck, M.V. Stoitsov and S. Wild, {\it Phys. Rev. C}, {\bf 85}, 024304(2012).
\bibitem{dddi} R.R. Chasman, {\it Phys. Rev. C}, {\bf 14}, 1935(1976).
\bibitem{dddi1} J.Terasaki, P.H. Heenen, P. Bonche, J. Dobaczewski and H. Flocard, {\it Nucl. Phys. A}, {\bf 593}, 1(1995). 
\bibitem{mix} J. Dobaczewski, W. Nazarewicz and M.V. Stoitsov, {\it Eur. Phys. J. A}, {\bf 15}, 21(2002).
\bibitem{sat} J.Terasaki, H. Flocard, P.H. Heenen and P. Bonche, {\it  Nucl. Phys. A}, {\bf 621}, 706(1997).
\bibitem{rad} N. Schunck and J. L. Edigo,  {\it Phys. Rev. C}, {\bf 78}, 064305(2008). 
\bibitem{miz} S. Mizutori, J. Dobaczewski, G. A. Lalazissis, W. Nazarewicz, and P.-G. Reinhard, {\it Phys. Rev. C}, {\bf 61}, 044326(2000).


\bibitem{constr} A. Staszczak, M.Stoitsov, A. Baran, and W. Nazarewicz, {\it Eur. Phys. J. A}, {\bf 46}, 85(2010).
\bibitem{ame2016} M. Wang, G. Audi, F. G. Kondev, W. J. Huang, S. Naimi and X. Xu {\it Chin. Phys. C}, {\bf 41}, 030003(2017).
\bibitem{magic} M. Bhuyan and S. K. Patra, {\it Pramana J. Phys.}, {\bf 82}, 851(2014).

\end{thebibliography}
\end{document}